\documentclass[twocolumnm,superscriptaddress,altaffilletter,prd]{article}

\usepackage[english]{babel}
\usepackage[utf8]{inputenc}
\usepackage{amssymb}
\usepackage{amsmath}
\usepackage{graphicx}
\usepackage[colorinlistoftodos,prependcaption,textsize=tiny]{todonotes}
\usepackage{geometry}
\usepackage{xfrac}
\usepackage{url}
\usepackage{upgreek}
\usepackage[normalem]{ulem}
\usepackage[ruled,vlined]{algorithm2e}
\usepackage{multirow}
\usepackage{dcolumn}
\usepackage{siunitx}

\usepackage[toc,page]{appendix}

\usepackage{diagbox} % for the slash in the table header

\usepackage{epigraph}
\usepackage[makeroom]{cancel}
\usepackage{multicol}
\usepackage[multiple]{footmisc}
\usepackage{bbm} % used it to get access to \mathbbm{1} to represent unitary matrices, since \mathbb does not act on numerals.
\usepackage{physics} % for bracket notation
\usepackage{doi}

\usepackage{longtable}
\usepackage{bm}
\usepackage[flushleft]{threeparttable}
\usepackage{nth}

\usepackage[switch]{lineno}
\usepackage[italic]{heppennames}
\usepackage{fixltx2e}

\usepackage{caption}
\usepackage{subcaption}

\usepackage{algorithmic}
\usepackage{cite}

\usepackage{hyphenat}
\modulolinenumbers[5]

\usepackage{array}

\usepackage{authblk}

\newcommand{\ie}{i.e.}
\newcommand{\eg}{e.g.}

\usepackage{xargs}

\usepackage{microtype}

\begin{document}

\title{A machine learning-based methodology for pulse classification in dual-phase xenon time projection chambers}

\author[1,*]{P. Brás}
\author[1]{F. Neves}
\author[1]{A. Lindote}
\author[2]{A. Cottle}
\author[1]{R. Cabrita}
\author[1]{E. Lopez Asamar}
\author[1]{G. Pereira}
\author[1]{C. Silva}
\author[1]{V. Solovov}
\author[1]{M. I. Lopes}

%\affil[1]{\textit{University of Coimbra, 3004-516 Coimbra, Portugal}}
\affil[1]{\small\textit{Laboratório de Instrumentação e Física Experimental de Partículas (LIP), University of Coimbra, P-3004 516 Coimbra, Portugal}}
\affil[2]{\small\textit{University of Oxford, Department of Physics, Oxford OX1 3RH, UK}}
\affil[*]{\small\textit{corresponding author - pbras@coimbra.lip.pt}}

\date{\today}

%%%%%%%%%%%%%%%%%%%%%%%%%%%%%%%%%%%%%%%%%%%%%%%%%%%%%%
\twocolumn[
  \begin{@twocolumnfalse}
    \maketitle
\begin{abstract}
% the major questions or hypothesis
Machine learning techniques are now well established in experimental particle physics, allowing detector data to be analysed in new and unique ways. The identification of signals in particle observatories is an essential data processing task that can potentially be improved using such methods. 
%the basic approach to answering that question
This paper aims at exploring the benefits that a dedicated machine learning approach might provide to the classification of signals in dual-phase noble gas time projection chambers. A full methodology is presented, from exploratory data analysis using Gaussian mixture models and feature importance ranking to the construction of dedicated predictive models based on standard implementations of neural networks and random forests, validated using unlabelled simulated data from the LZ experiment as a proxy to real data. 
% key results and data
The global classification accuracy of the predictive models developed in this work is estimated to be $>$99.0\%, which is an improvement over conventional algorithms tested with the same data.
% a brief conclusion about what you discovered
The results from the clustering analysis were also used to identify anomalies in the data caused by miscalculated signal properties, showing that this methodology can also be used for data monitoring.

%\paragraph{Keywords:} TPC, Xenon, Classification, Machine Learning, Dark Matter Search, Rare Event Search
\paragraph{}~
\end{abstract}
  \end{@twocolumnfalse}
]

%%%%%%%%%%%%%%%%%%%%%%%%%%%%%%%%%%%%%%%%%%%%%%%%%%%%%%
\section{Introduction}
\label{sec:intro}

Dual-phase noble element time projection chambers (TPCs) are excellent rare event observatories due to their low background, low energy threshold, good energy and position resolutions, and ability to scale their target mass \cite{Chepel:2012sj}. These detectors have a long history in direct searches for dark matter (DM) in the form of Weakly Interacting Massive Particles (WIMPs) \cite{ZEPLINIII, Akerib:2016vxi, Agnes:2018ves,  Aprile:2018dbl, Wang:2020coa} and new, multi-tonne scale detectors using this technology are already or will soon start collecting data in the near future \cite{Aalbers:2016jon,Akerib:2019fml,Aprile:2020vtw}. 
A dual-phase noble element TPC consists of a scintillating liquid target volume (usually argon or xenon) in equilibrium with a gaseous layer on top, both observed by arrays of light sensors placed at the top of the gas phase and/or at the bottom of the liquid phase.
An interaction with the target material in the TPC will excite and ionize some atoms at the interaction site, producing both scintillation light and ionization charge \cite{Chepel:2012sj}. The scintillation light is promptly detected by the light sensors as the primary signal (S1). An electric field across the liquid phase drifts the ionization electrons that did not recombine towards the liquid-gas interface, where they are extracted to the gas phase and accelerated in a stronger field to produce a larger electroluminescence light signal (S2). 
The distribution of the S2 light over the light sensor arrays is used to reconstruct the position of the initial interaction in the horizontal plane, while the time between the S1 and S2 signals indicates the depth of the interaction. Furthermore, the relative size of these signals can be used to distinguish between electron recoils (ER) and nuclear recoils (NR) and infer the nature of the interacting particle \cite{LUX:2020car}. 

In order to fully characterize an interaction in both position and energy, all signals resulting from the different processes in the detector must be correctly identified. 
This is usually performed by a specialized classification algorithm. 
Such classifiers need not only to correctly identify the main signals (S1 and S2) that characterize an event in a dual-phase TPC but also avoid misclassifying spurious signals as relevant ones. A high efficiency classifier is imperative in order to ensure that no event is misreconstructed and that rare signal events are accurately identified.
Dual-phase noble element TPCs share the same overall pulse shape characteristics for a wide range of detector parameters, like electric field strength, light collection efficiency or the thermodynamic parameters of xenon/argon \cite{Gonzalez-Diaz:2017gxo}, and thus also share many of the same challenges related to signal identification. 

Several techniques can be deployed to classify the signals recorded in these detectors, from simple, human-built, heuristic methods like decision trees \cite{Akerib:2017vbi,XENON:2017lvq} to advanced data analysis techniques such as machine learning (ML). 
It has been demonstrated that a classification efficiency for S1 and S2 signals above 90\% is achievable using conventional heuristic methods \cite{Aprile:2018dbl,Wang:2020coa}, with some signal-level processing methods showing efficiencies $>$98\% and $>$99\% for small S1 and S2 signals, respectively \cite{Akerib:2013tjd}. 
The implementation of classification methods based on ML could improve this signal identification efficiency while also providing important information for the identification of spurious pulses. These methods have the advantage of handling high dimensional data more efficiently than conventional methods and can uncover unique and insightful information about intrinsic properties of the data. 
This work will explore a ML-based methodology for the development of high-accuracy and minimally-biased pulse classification tools tailored for dual-phase noble element TPCs. 

%-----------------------------------------------------
% MOTIVATION
There is an underlying risk in training a ML classifier model with non-representative data, most notably if the training is performed in a supervised manner. Supervised learning methods often resort to simulated data for training, and their results are therefore dependent on the accuracy of those simulations. This motivates the usage of unsupervised learning methods that do not rely on approximated models of the real data. 
In these approaches, the data features used for training are obtained directly from the target dataset, guaranteeing that all the available information and any possible data trends are present with no underlying hidden biases. 

A particularly convenient set of tools used to process unlabelled datasets are clustering algorithms. Clustering analysis is very often combined with classification efforts because it naturally partitions the data based on its intrinsic properties in a robust, minimally-biased way \cite{ElementsStatisticalLearning}.
In this work, a clustering analysis using Gaussian mixture models (GMMs) \cite{bishop2006pattern} is used to build two distinct predictive models designed for classification, those being random forests (RFs) \cite{RandomDecisionForestsTinKamHo,RandomForestsBreiman} and an ensemble of neural networks (NN) \cite{NNsAndDeepLearningAggarwal2018}. 
This clustering method was chosen for its ability to handle large density inhomogeneities in the data \cite{ElementsStatisticalLearning}.
The predictive models were selected due to their simplicity, robustness and excellent performance as classifiers, as well as their availability across many different ML software packages and implementations. The RF model was also selected due to its ability to extract information about the best discriminant features in the data \cite{ElementsStatisticalLearning}.

%-----------------------------------------------------
Section \ref{sec:dataprocessing} will give an overview of the data used in this analysis (unlabelled simulated data from the LUX-ZEPLIN (LZ) experiment \cite{Akerib:2019fml,Akerib:2020ewf}). Section \ref{sec:method} will present the methodology developed for processing the data and constructing classification tools based on ML: a clustering analysis is presented in Section \ref{subsec:MLClustering}, that will then be the basis for the development of predictive models based on RFs presented in Section \ref{subsec:DTClass} and a NN ensemble model presented in Section \ref{subsec:TriNet}. A discussion of the results and main issues can be found in Section \ref{subsec:Discussion}. Section \ref{sec:concl} will present some final remarks regarding the different models and an overview of the methodology developed in this work.

%%%%%%%%%%%%%%%%%%%%%%%%%%%%%%%%%%%%%%%%%%%%%%%%%%%%%%
\section{Overview of the Data}
\label{sec:dataprocessing}

This work was developed using LZ simulated data \cite{Akerib:2020ewf} from a simulated dataset created on October 2018 by the collaboration to validate its analysis tools. 
Simulated LZ data was chosen in order to keep this work relevant for current and next generation experiments searching for rare events over a wide range of energies. This data was also readily available and extensively tested. 
To ensure that the simulated data is analogous to real data, Monte Carlo truth information was not used for this study. 

LZ will use a 7~tonne xenon TPC observed by 494 photomultiplier tubes (PMTs) distributed between two arrays (253 at the top and 241 at the bottom) \cite{PhysRevD.101.052002}. 
An event is comprised of digitized waveforms originating from the readout of the TPC PMTs. 
The waveforms are digitized at 100~M samples per second and are only recorded when the voltage response of the PMT is above a certain threshold \cite{Akerib:2019fml}. 
These are then combined across all PMT channels to produce a summed waveform, which will contain several distinguishable structures that can be isolated in time and are expected to correlate to particular signals in the TPC. 
Each such structure will be referred to as a pulse. 
The event data is processed in a typical modular analysis approach: after some initial time calibrations and baseline corrections, dedicated pulse finding algorithms isolate individual pulses in time, followed by parameterization algorithms that calculate the properties of these pulses, which are then used to identify their origin by the classifier algorithms. After this pulse-level analysis, other dedicated algorithms use the full context of the event to characterize the interaction in the detector. 

Errors along the data processing chain, from the electronics readout to the pulse-level analysis algorithms, can produce spurious pulses that contaminate the data. The correct identification and characterization of these spurious pulses is critical for the physics analysis, as well as to mitigate them at their origin.

%%%%%%%%%%%%%%%%%%%%%%%%%%%%%%%%%%%%%%%%%%%%%%%%%%%%%%
\subsection{Signals in a Dual-Phase Noble Element TPC}
\label{subsec:Signals}

\begin{figure*}[!ht]
    \centering
    \begin{subfigure}[b]{0.49\textwidth}
        \centering
        \includegraphics[width=\textwidth]{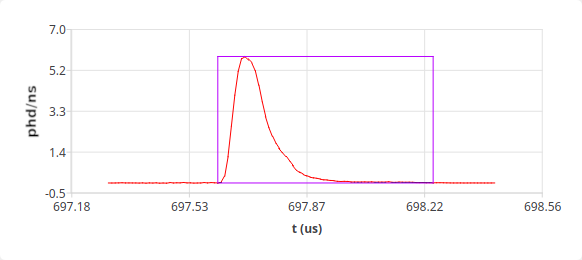}
        \caption{Typical shape of an S1 pulse (purple).}
        \label{fig:S1_example}
    \end{subfigure}
    \hfill
    \begin{subfigure}[b]{0.49\textwidth}
        \centering
        \includegraphics[width=\textwidth]{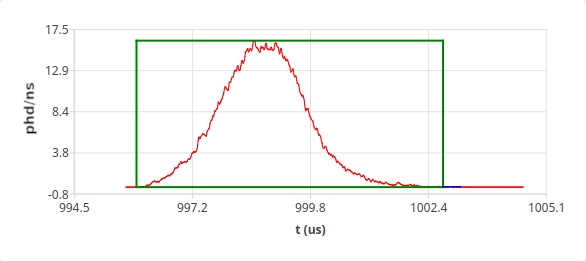}
        \caption{Typical shape of a low-energy S2 pulse.}
        \label{fig:S2_example}
    \end{subfigure}
    \hfill
    \begin{subfigure}[b]{0.49\textwidth}
        \centering
        \includegraphics[width=\textwidth]{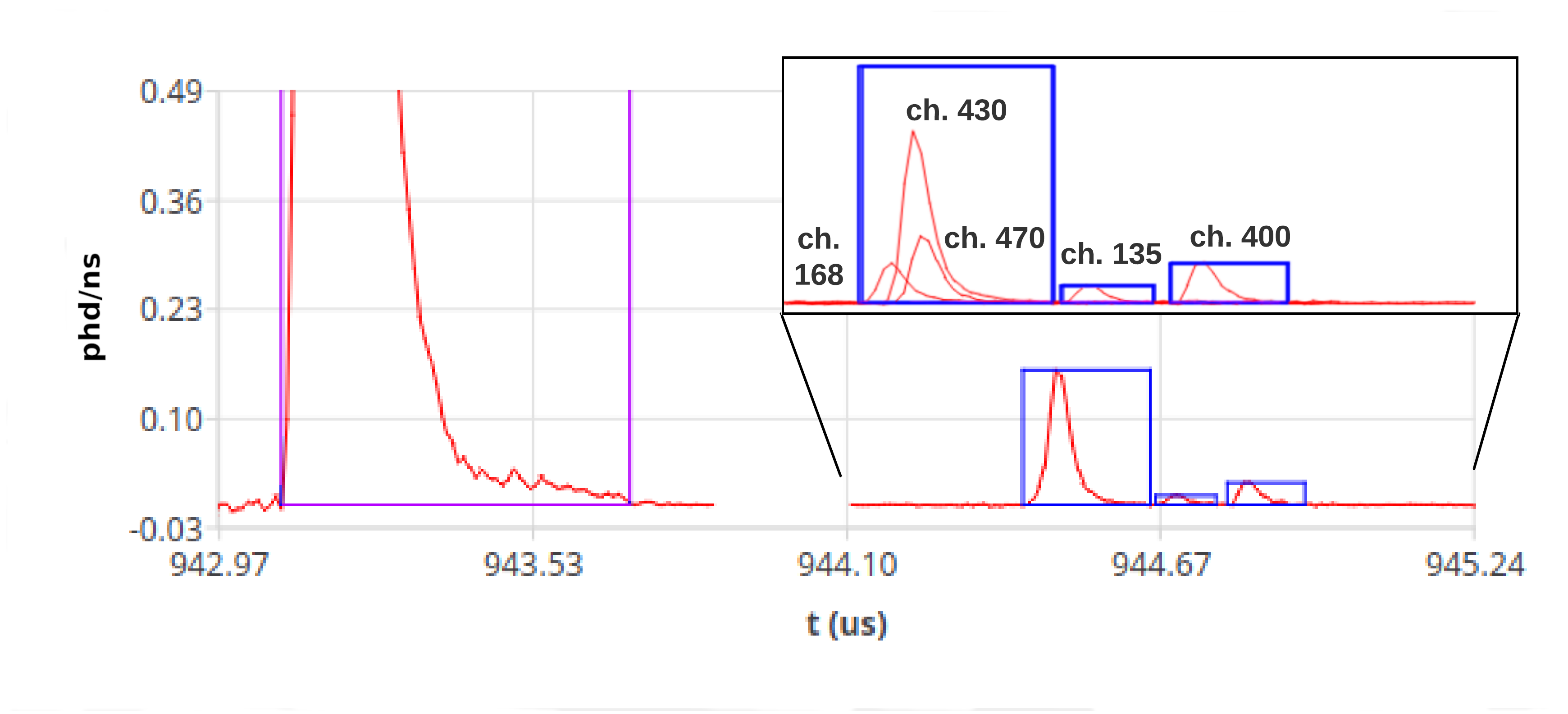}
        \caption{S1 pulse followed by PMT afterpulsing (AP), with individual PMT channels identified in the plot inlay.}
        \label{fig:AP_example}
    \end{subfigure}
    \hfill
    \begin{subfigure}[b]{0.49\textwidth}
        \centering
        \includegraphics[width=\textwidth]{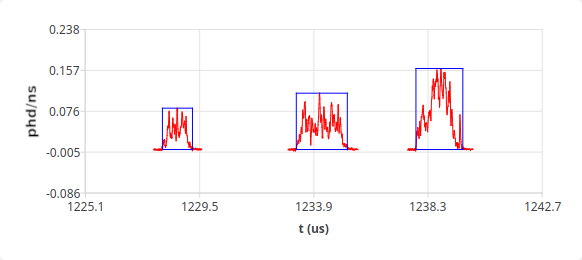}
        \caption{A typical SE pulse (left) followed by two SE pileup examples: merged (middle) and summed (right).}
        \label{fig:SEpileup_example}
    \end{subfigure}
    \hfill
    \begin{subfigure}[b]{0.49\textwidth}
        \centering
        \includegraphics[width=\textwidth]{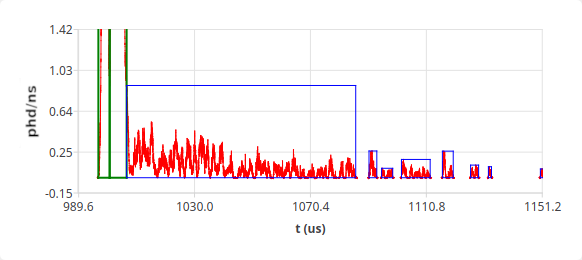}
        \caption{S2 tails and e-trains following two large S2 signals which extend beyond the vertical range (green).}
        \label{fig:S2tails_example}
    \end{subfigure}
    \hfill
    \begin{subfigure}[b]{0.49\textwidth}
        \centering
        \includegraphics[width=\textwidth]{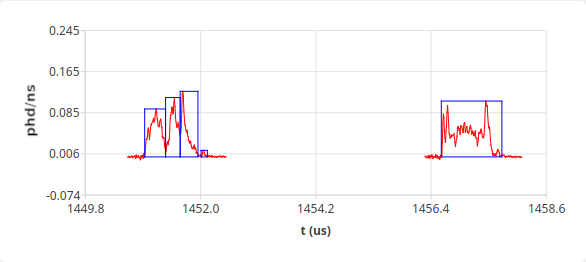}
        \caption{Example of a SE pulse split by the pulse finding algorithm (left) and a correctly identified SE pulse (right).}
        \label{fig:SEsplit_example}
    \end{subfigure}
    \hfill
    \caption{
    \label{fig:LZapPulsesExample}
    Examples of typical pulses expected in dual-phase noble element TPCs, obtained from LZ simulated data. Pulse amplitudes are converted to units of photons detected (phd) per nanosecond. The coloured boxes represent the pulse boundaries calculated by a simple pulse finding algorithm, the purple and green boxes mark the main S1 and S2 pulses of an event, respectively. The splitting of single electron (SE) pulses, like the example presented in Figure \ref{fig:SEsplit_example}, is particularly challenging to the classification, since these spurious pulses have characteristics that are very similar to S1 pulses and PMT afterpulsing (AP).}
\end{figure*}

Figure \ref{fig:LZapPulsesExample} displays some examples of pulses in the dataset used that are expected to be found in a dual-phase noble element TPC. 
The main pulse types are the aforementioned S1 and S2, that are represented in Figures \ref{fig:S1_example} and \ref{fig:S2_example}, respectively. 

%----------------------- explain s1
In the presence of an electric field, most of the scintillation light is produced by the decay of xenon excimers, which have fast decay times of 2.2~ns and 27~ns for transitions from singlet or triplet electronic excited states to the ground state, respectively \cite{Chepel:2012sj}. 
Due to a combination of the pulse response of the PMT amplifier chain and S1 photon flight time, which can become greater than the decay times of the xenon excimers in large TPCs \cite{Moongweluwan:2015nks}, a typical S1 pulse has a length in time of the order of 100~ns FWHM, rising quickly ($\sim$50~ns) and falling exponentially ($\sim$500~ns). The S1 signals will also have, on average, larger amplitudes in the bottom PMT channels due to the internal reflection of scintillation in the liquid-gas interface. 

%----------------------- explain s2
The S2 signal is proportional to the number of electrons extracted into the gas phase, with each electron producing hundreds of scintillation photons \cite{Chepel:2012sj}. The shape in time of the S2 signal is dictated by the electron transit time across the gas gap in the extraction region and by the charge distribution of the electron cloud drifted from the interaction site to the liquid-gas interface \cite{LZTDR:2016}. The drifting electrons will experience diffusion (in both the longitudinal and transverse directions) along the travel path, making the S2 signals from interactions deeper in the TPC wider than those from interactions closer to the liquid-gas boundary. These effects result in S2 pulses with a length in time of the order of $\sim$2$~\mu$s~FWHM, considering an extraction field of $\gtrsim$10~kV/cm and a gas gap of 8~mm (electron transit time of 1.2~$\mu$s) \cite{LZTDR:2016,Akerib:2019fml}. 
Due to electron diffusion, the S2 pulses tend to have an almost symmetrical shape in time, akin to a normal distribution. Most of the S2 light is detected by the top PMT array due to its proximity to the liquid boundary and extraction region.

%----------------------- explain s1 low photon limit
A pulse that results from the detection of a single photon by a PMT is called a single photoelectron (SPE). The smallest S1 pulses are composed of SPEs in a small set of PMT channels within a short time window. Random coincidence of PMT dark counts (spontaneous and spurious PMT pulses inherently indistinguishable from SPEs) can mimic small S1 pulses. To minimize this effect, a channel coincidence (\ie, the number of PMT channels that record signal within the time window of a pulse) of 3 or more is required for a pulse to be considered an S1 in LZ (3-fold coincidence) \cite{PhysRevD.101.052002}. In order to preserve potential smaller S1 signals in low-energy analyses, the channel coincidence requirement for S1 pulses was reduced to 2-fold in this work. This wider acceptance of pulses with lower channel coincidence can be reverted at a later analysis stage, if needed.

%----------------------- explain afterpulsing
PMT afterpulses (APs) are typically caused by residual gas ionized by the electrons accelerated in the charge multiplication stages of the PMT. These ions are then drifted back to the photocathode where they release more electrons and produce a delayed signal at characteristic times \cite{DEAP:2017fgw,LopezParedes:2018kzu}. 
Figure \ref{fig:AP_example} shows an S1 pulse followed by five APs. 
The figure inlay shows the five channels that produced afterpulsing. The APs in channels 168, 430 and 470 are overlapping in time and were merged into a single pulse, resulting in only three delayed pulses being observed in the summed response of all PMT channels. The first of these three pulses has coincidence 3 and resembles an S1 pulse in shape, making it indistinguishable from a true S1 pulse without the context of the full event.

%----------------------- explain s2 low electron limit
An example of a signal generated by the extraction of a single electron (SE) from the liquid into the gas phase is presented in Figure \ref{fig:SEpileup_example} (left-most pulse) and in Figure \ref{fig:SEsplit_example} (see discussion below for more details on the latter example). An S2 pulse is, in its essence, the overlap of several SE signals. 
The shape of SE pulses display great variability, with characteristic spike structures caused by the random nature of the detection of electroluminescence photons.
As mentioned previously, the extraction of a single electron can produce hundreds of photons \cite{Chepel:2012sj}, with the average number of photons detected per extracted electron depending on the fixed strength of the extraction field, photon yield of the gas and light collection efficiency of the detector. Therefore, SE pulses have a very characteristic size that is used to calibrate the response of the detector \cite{Akerib:2019fml}. 
Much like the channel coincidence criteria for S1 pulses, S2 pulses usually need to have areas several times larger than the average SE area in order to be considered valid, excluding false S2 pulses caused by random SE pileup. However, in order to preserve the classification of the smallest S2 pulses, the pileup of two or more SE pulses will be considered a valid S2 pulse in this work.

%----------------------- explain s2 tails and such
S2 pulses are generally followed by a tail of delayed electron emissions that can result in SE pileup and thus mimic an S2 signal, as seen in Figure \ref{fig:S2tails_example}. These delayed emissions typically produce an immediate continuous signal within some tens of $\mu$s after the S2 signal (dubbed ``S2 tail'') and a disperse trail of SE pulses that can last several ms after the S2 (``e-trains'') \cite{Sorensen:2017kpl,Akerib:2020jud}. These emission phenomena tend to scale with the size of the main S2 signal. The S2 tail is most likely the result of photoionization of impurities in the liquid bulk caused by xenon luminescence, and by photoelectric effect on the field grids. The e-trains and other SE delayed emissions with longer time scales are likely produced by the spontaneous emission of electrons trapped at the liquid boundary and by the capture and subsequent release of drifting electrons by impurities\cite{Sorensen:2017kpl,Akerib:2020jud}.

%----------------------- explain spurious pulses
Spurious pulses, \ie, pulses that do not directly correlate to real light-yielding processes in the TPC or that present unusual or unexpected properties, are the result of errors at any stage of data handling, from the electronics readout to the pulse-level analysis algorithms.
These pulses often display similarities with either S1 or S2 pulses, contaminating the data and leading to misclassification errors. 
Figure \ref{fig:SEsplit_example} shows two SE pulses, with the leftmost being incorrectly split into several non-physical substructures by the pulse finding algorithm. 
These individual structures resemble S1 pulses in both shape and timing, and constitute a significant challenge for pulse-level classification in this dataset.

%----------------------- introduce the main classes here
The main pulse classes that are usually considered in the low-level analysis of dual-phase noble-element TPCs are the aforementioned S1, S2, SE and SPE classes. 
Pulses that do not belong to any of the previously mentioned classes or that have non-physical properties (\ie, spurious pulses) are labelled as ``Other'' in this work.

%%%%%%%%%%%%%%%%%%%%%%%%%%%%%%%%%%%%%%%%%%%%%%%%%%%%%%
\subsection{Pulse Features and Data Preprocessing}
\label{subsec:DadasetOverview}

%----------------------- Discussion of the RQs
Pulses are identified by the classification algorithms based on their characteristic geometrical features. 
Some analyses use the raw shape of the pulse directly to identify the pulse (\eg, using convolutional neural networks \cite{Griffiths:2018zde} or deep learning \cite{Holl:2019xtt}) but typically an intermediate parameterization step is used to calculate relevant shape-related quantities, such as integrated areas, lengths, fit parameters and other detector-specific traits, that are then used by the classification algorithms. This work will follow the latter approach. 
Of the available pulse features in the LZ dataset used, dubbed ``reduced quantities'' (RQs), 13 were selected for this analysis and are summarized in Table \ref{tab:RQs}.  
These RQs were selected based on past experience on heuristic pulse classification algorithms developed for the LUX and LZ experiments. 

\begin{table*}[!ht]
    \centering
    \caption{Some of the pulse-level RQs available in LZ simulated data chosen for the classification analysis in this work.}
    \label{tab:RQs}
    \begin{tabular}[c]{ >{\arraybackslash} m{0.13\textwidth} | >{\arraybackslash} m{0.06\textwidth} | p{0.65\textwidth} }
        Name [unit] & Type & Description \\
    	\hline
	    \textit{pA} [phd] & \textit{float} &  Total integrated area from the start to the end of the pulse. \\
	    \textit{pH} [phd/ns] & \textit{float} & Pulse maximum amplitude. \\
	    \textit{pHT} [ns] & \textit{int} & Time at which the pulse reaches maximum amplitude. \\
	    \textit{pL} [ns] & \textit{int} & Time difference between start and end of the pulse. \\
    	\textit{pL90} [ns] & \textit{int} & Pulse length time at 90\% area, from 5\% to 95\% integrated area time. \\
    	\textit{pRMSW} [ns] & \textit{int} & Pulse root mean square (RMS) width. \\ 
    	\textit{pF50} & \textit{float} & Fraction of the pulse area integrated in a 50~ns time window starting 10~ns before the 5\% integrated area time. \\
    	\textit{pF100} & \textit{float} & Same as \textit{pF50} but for a 100~ns integration window. \\
    	\textit{pF200} & \textit{float} & Same as \textit{pF50} but for a 200~ns integration window. \\
    	\textit{pF1k} & \textit{float} & Same as \textit{pF50} but for a 1~$\mu$s integration window. \\
    	\textit{TBA} & \textit{float} & Top-bottom asymmetry: Difference between the top PMT area fraction and bottom PMT area fraction. \\
    	\textit{pHTL} & \textit{float} & Ratio between \textit{pHT} and \textit{pL}. \\
    	\textit{coincidence} & \textit{int} & Number of PMT channels that record signal within pulse boundaries. \\
    	\hline
    \end{tabular}
\end{table*}

A pre-selection of pulses was performed to ensure the quality of the dataset before the classification analysis, excluding pulses with non-physical RQ values, \eg, negative total areas, height or length. 
Pulses with \textit{coincidence}$=$1 are also excluded \textit{a priori} from this analysis since these can be automatically identified as either SPE or Other pulses, both of which are not critical for the description of the event.
However, the value of the \textit{coincidence} RQ is often overestimated in this dataset due to accidental partial overlap of pulses across different channels, especially for SPE pulses and afterpulsing. 
Even though these pulses are expected to be measured in a single PMT channel, negligible contributions from baseline fluctuations or small portions of partially overlapping pulses in other channels within the boundaries of the pulse can lead to higher \textit{coincidence} RQ values. 
This results in some SPEs and APs having \textit{coincidence}~$> 1$ and thus are eligible to be classified as S1 pulses, contaminating the RQ dataset. 
This, however, does not affect the final physics analyses since these distinctions are intentionally delayed to a dedicated event identification algorithm that uses the context provided by the remaining pulses in the event. 
This separation between pulse-level and event-level processing allows for the deployment of algorithms that are highly specialized for each task.

\begin{figure*}[!ht]
    \centering
    \includegraphics[width=0.45\textwidth]{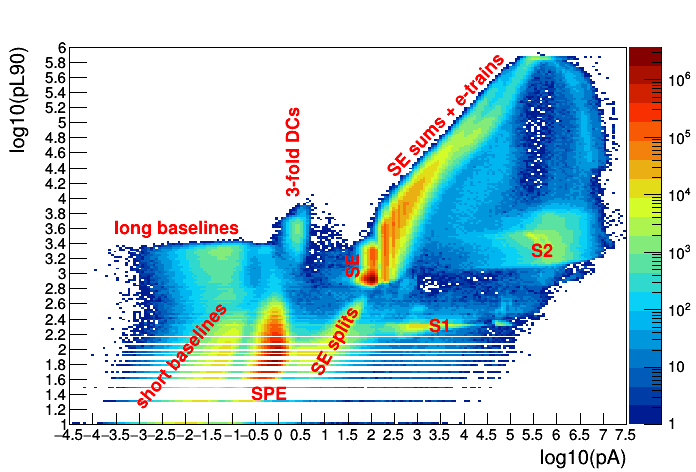}
    \includegraphics[width=0.45\textwidth]{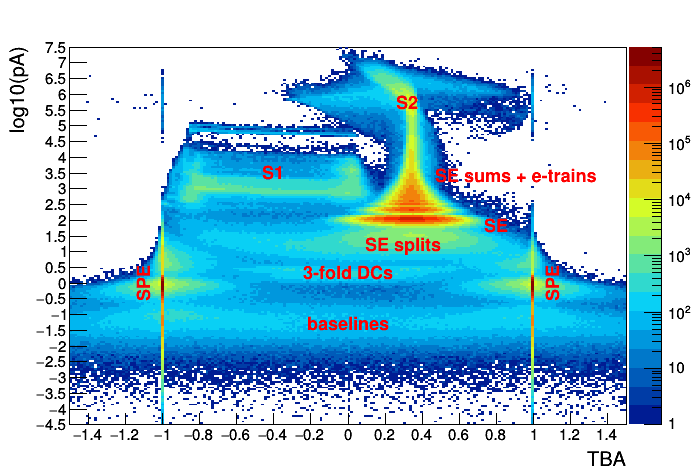}
    \caption{Marginal distributions of the main pulse populations in pulse area (\textit{pA}) vs pulse length at 90\% area (\textit{pL90}) space (left) and top-bottom asymmetry (\textit{TBA}) vs \textit{pA} space (right). Not all isolated populations are labelled, only the cases presented in Figure \ref{fig:LZapPulsesExample} have been highlighted. The populations labelled ``baselines'' consist of random noise that is isolated by the pulse finding algorithm. The population labelled ``3-fold DC'' contains spurious pulses caused by random coincidences of dark counts in 3 PMT channels.}
    \label{fig:LZap_populations}
\end{figure*}

After the quality selections mentioned above, a total of 10$^{6}$~pulses are randomly selected from the remaining. The full dataset corresponds to an average of 25400 distinct events, corresponding to roughly 10 minutes and 35 seconds of exposure considering the expected average background event rate of 40~Hz for LZ \cite{Akerib:2019fml}. 
Figure \ref{fig:LZap_populations} shows a rough labelling of the main pulse populations in the dataset, obtained using traditional pulse identification methods and visual inspection. The goal of Figure \ref{fig:LZap_populations} is to act as a reference for the upcoming results in Sections \ref{subsec:MLClustering}, \ref{subsec:DTClass} and \ref{subsec:TriNet}. The populations are not fully separable in any 2-dimensional representation of pulse features but their general distributions can be intuitively inferred.

%%%%%%%%%%%%%%%%%%%%%%%%%%%%%%%%%%%%%%%%%%%%%%%%%%%%%%
%%%%%%%%%%%%%%%%%%%%%%%%%%%%%%%%%%%%%%%%%%%%%%%%%%%%%%
\section{Method}
\label{sec:method}

The following sections explain in detail the implementations and results of the ML methods explored in this analysis. A clustering analysis of the data is performed first, using Gaussian mixture models, and the results are then used to train two predictive models based in tree ensembles (RFs) and NNs to perform pulse classification and to extract useful information from the RQ dataset. 
These ML methods are extensively used in the physical sciences \cite{Carleo:2019ptp}. 
The potential of tree ensemble methods like RFs and boosted decision trees (BDTs) for data analysis in physics is well established, from searches for beyond the Standard Model physics at particle colliders \cite{{Bertone:2016mdy,Bertone:2017adx}} to dark matter direct detection experiments \cite{2016JCAP.05.019A,Wang:2020coa,LUX:2020yym}. 
 
Much like the tree ensemble methods, NNs are extensively used in dark matter searches at colliders \cite{Bertone:2017adx,Dey:2019lyr,Khosa:2019kxd,Shirobokov:2020tht}, direct detection experiments \cite{SuperCDMS:2015lke,PICO:2018fbf,Khosa:2019qgp}, cosmology and astrophysics surveys \cite{Brehmer:2019jyt,Alexander:2019puy,Escamilla-Rivera:2019hqt}, and in other rare event searches \cite{EXO:2018bpx,NEXT:2020jmz}. 
However, this work will take a different approach to the traditional implementation of a single feed-forward dense NN by implementing instead an ensemble of NN classifiers trained individually to classify each valid pulse class in the data in a One-vs-All configuration \cite{bishop2006pattern}. 
Ensembles of NNs have been around for at least three decades \cite{58871,ZHOU2002239,10.1007/978-3-540-74690-4_11,2019arXiv190405488T}, but the implementations are often based on generalizations of boosting or bagging strategies applied to NN models. 
In this work, the NN ensemble is instead composed of specialized binary classification models that work in parallel and whose results are combined to obtain a final classification.

%------------------------------------------------------
% The classification end-goal
Throughout this work the dataset will be partitioned into four distinct classes, corresponding to the class labels $$\mathcal{S}=\{s_1, s_2, s_3, s_4\}=\{\text{S1, S2, SE, Other}\}.$$ 
The S1 pulse class will include 2-fold coincidence pulses in an effort to preserve the classification efficiency for low-energy searches. 
The distinction between S2 and SE pulses, despite both having the same physical origin, is expected to help the classification algorithms distinguishing between larger S1 pulses and the plethora of S2-like pulses, such as S2 tails and SE pileup, during training. 
%------------------------------------------------------

%%%%%%%%%%%%%%%%%%%%%%%%%%%%%%%%%%%%%%%%%%%%%%%%%%%%%%
\subsection{Clustering Analysis}
\label{subsec:MLClustering}

GMMs are a multi-component probabilistic distribution where a finite ensemble of Gaussian distributions are assumed to originate the observed data \cite{bishop2006pattern}.
The GMM implementation of the public python package  \textit{scikit-learn} \cite{scikit-learn} was used for the cluster analysis\footnote{\url{https://gitlab.com/PauloBras/gmmclustering.git}}\footnote{\url{https://gitlab.com/PauloBras/mlforpc.git}} considering $K=67$~components and full covariance freedom between the different GMM components. The number of components of the mixture model was roughly estimated using an implementation of the kernel-based algorithm described in Reference \cite{RKHSClusteringClassifier}. This number was left purposefully larger than the number of categorical pulse classes being considered by the classifier module since each global pulse class contains several distinct pulse populations. This increases the efficiency of the clustering analysis by allowing a more ``fine-grained'' model that can account for the hidden stratification of the data \cite{sohoni2020no}. 
Also, over-partitioning allows for a deeper understanding of the differences between seemingly degenerate populations, possibly allowing for the detection of outlier populations or pathological structures that were not identified in the preceding steps. 
Furthermore, after the clustering analysis, the different components of the mixture model can be collapsed into the same categorical classes corresponding to the pulse types expected in the data. 
This is a form of hierarchical processing that greatly accelerates learning on the subsequent classification algorithms \cite{wolfgangIntroAI2017}. However, some of the choices made in the assignment of pulse labels to the GMM components may not be ideal and will introduce biases in the following processing steps. As with many aspects of cluster analysis, and ML in general, a heuristic approach is somewhat inevitable \cite{ElementsStatisticalLearning}.

\begin{figure*}[!ht]
   \centering
   \includegraphics[width=0.45\textwidth]{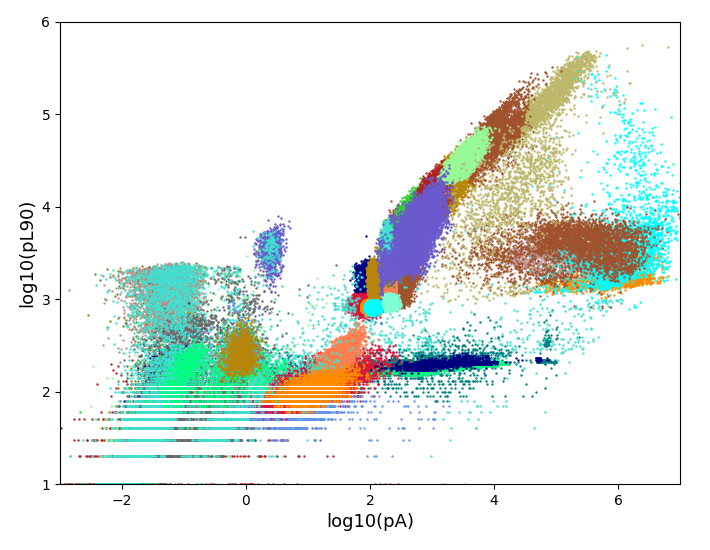}
   \includegraphics[width=0.45\textwidth]{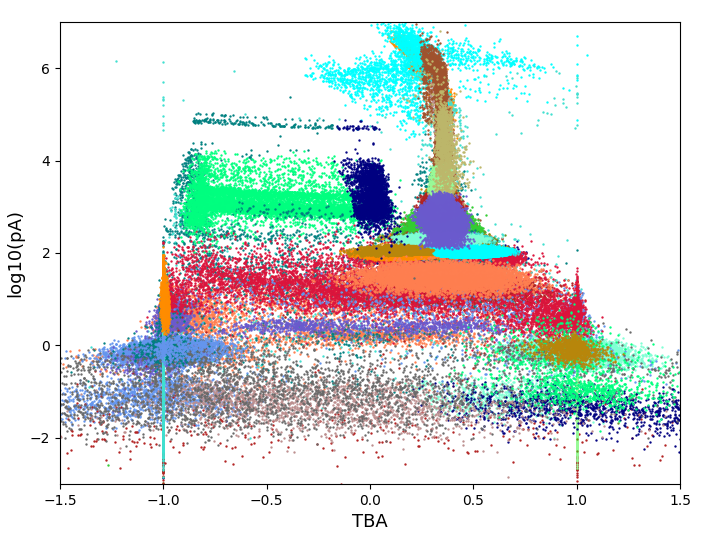}
   \caption{Scatter plots of the GMM components for pulse area (\textit{pA}) vs pulse length (\textit{pL90}) RQs (left) and top-bottom asymmetry (\textit{TBA}) vs pulse area RQs (right). The colors associated to each GMM component are cyclical and repeat for some components. The size and density of the 67 components vary significantly.}
   \label{fig:GMM_bases_distribution}
\end{figure*}
Figure \ref{fig:GMM_bases_distribution} displays the results of the GMM clustering of the RQ dataset, considering $K=67$ components and all pulse RQs in Table \ref{tab:RQs}. Despite the probabilistic nature of the GMM algorithm, each pulse is categorically associated to the Gaussian component with the highest likelihood of having generated it \cite{bishop2006pattern}.
Every GMM component is checked for spurious features in these and other marginal distributions and any strange population is handscanned, \ie, a sample of pulse waveforms are analysed by eye, to determine their constituents. The first noticeable feature is the cyan population of S2-like pulses at high \textit{pA} and trailing to higher \textit{pL90} values that clearly displays a pathological behaviour of \textit{TBA} (range outliers), with fluctuations that are larger than expected for regular S2 pulses (see Figure \ref{fig:LZap_populations}). Some SE signals that are followed by either an afterpulsing or a coincident dark count seem to be similar enough to small S2 pulses in the RQ space, and thus the GMM model has produced some components with a small mixture of SE and S2 pulses. This, however, is inconsequential to the physics analysis since S2 and SE pulses can be discriminated by area alone in a subsequent processing step.

Each GMM component is now assigned a categorical pulse class based on the type of pulses that it contains. The contents of each component are inferred by carefully characterizing their parameter distributions and by handscanning a larger sample of pulses. 
Two GMM components were identified as containing most of the SE split population with low contamination from other pulse species. However, handscans of other GMM components seem to indicate that the clustering analysis cannot fully resolve afterpulsing and the contaminant SPE pulses from other spurious pulses like SE splits or baseline fluctuations. The components with noticeable contamination are assigned to the categorical pulse class in the majority. 
A discussion of the effect that these results and choices have in the training of the predictive models is provided in \mbox{Section \ref{subsec:Discussion}}.

A rough classification of the data has already been performed by identifying the contents of each GMM component. 
Even though mixture models can be used as predictive models, alternative ML methods such as tree ensembles and neural networks are often better at generalizing and handling novel data. 
In the absence of labelled data, the results from the GMM clustering analysis can be used as a prior estimate of the pulse classes and were used to train the predictive models described in Sections \ref{subsec:DTClass} and \ref{subsec:TriNet}. 
The accuracy of these models will be calculated using the results obtained in this Section as targets, together with careful monitoring of their results via data handscans.

%%%%%%%%%%%%%%%%%%%%%%%%%%%%%%%%%%%%%%%%%%%%%%%%%%%%%%%%%%%%%
\subsection{The RFClassifier Pulse Classification Tool}
\label{subsec:DTClass}

The simplicity and robustness of random forests motivated the development of the RFClassifier\footnote{\url{https://gitlab.com/PauloBras/rfclassifier.git}}\footnote{\url{https://gitlab.com/PauloBras/mlforpc.git}} pulse classification tool. This tool has been developed using the \textit{scikit-learn} implementation of the \textit{RandomForestClassifier} model \cite{scikit-learn}. The methodology followed here aims to provide a deeper understanding of how to best separate bad pulse populations overlapping with the main populations and to determine which RQs, parametric thresholds, and sequences of selection criteria yield the most efficient partitioning of the dataset (\ie, feature importance ranking) \cite{ElementsStatisticalLearning,RandomForestsBreiman}.

The RQs selected for benchmarking the model were \textit{pA}, \textit{pH}, \textit{pL90}, \textit{pRMSW}, \textit{pF50}, \textit{pF100},\textit{pF200}, \textit{pF1k},\textit{TBA} and \textit{pHTL}. The \textit{coincidence} RQ was not selected due to being miscalculated for smaller pulses, as explained in Section \ref{subsec:DadasetOverview}. The \textit{pL90} RQ provides a better estimation of the length of a pulse compared to \textit{pL}, with the latter presenting more variance for smaller pulses and being only used to calculate the \textit{pHTL} RQ. Despite being highly correlated, the four prompt fraction RQs (\textit{pF50}, \textit{pF100}, \textit{pF200} and \textit{pF1k}) were included in the input data to determine which one has the strongest discriminant power among them.

The results obtained in the clustering analysis in Section \ref{subsec:MLClustering} are used to train and benchmark the RFClassifier model. No additional selection of data was performed and all classes are considered to have the same importance. It is worth noting that some classes are more common than others in this dataset, namely SE pulses. Bootstrap aggregating (bagging) may induce some class bias if some classes are more represented than others in a multi-class dataset \cite{ElementsStatisticalLearning,RandomForestsBreiman}. This is especially damaging if bootstrapping is done without replacement, which is the case here. However, the asymmetry on the abundances of the different species is not too severe and all classes are assumed to be sufficiently well represented, with the least prevalent class (S1 pulses) corresponding to around 10\% of all pulses in the dataset. 

The number of trees in the model, their individual depth, and the minimal sample size to allow the data to be split, were estimated recursively by monitoring the performance of the model. The final model is composed of 101 learners with no limitation of growth and minimum number of samples to split a branch set at 2. The RQ dataset was divided into a training set and test set with 80--20\% splitting ratio.
\begin{figure*}[!ht]
   \centering
   \includegraphics[width=0.45\textwidth]{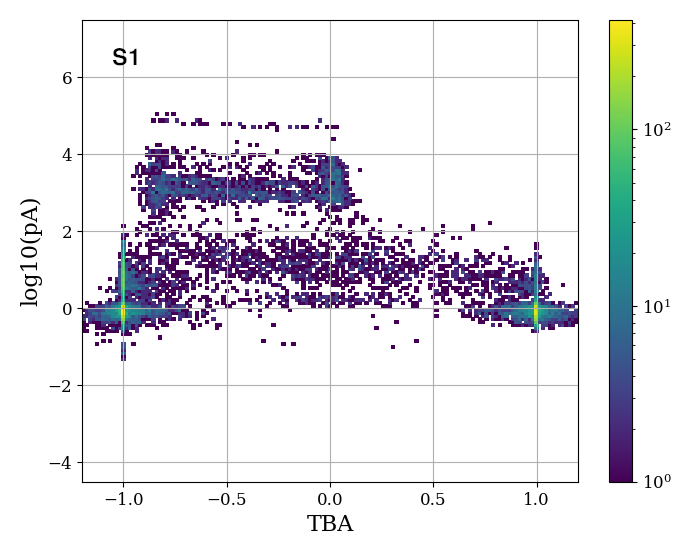}
   \includegraphics[width=0.45\textwidth]{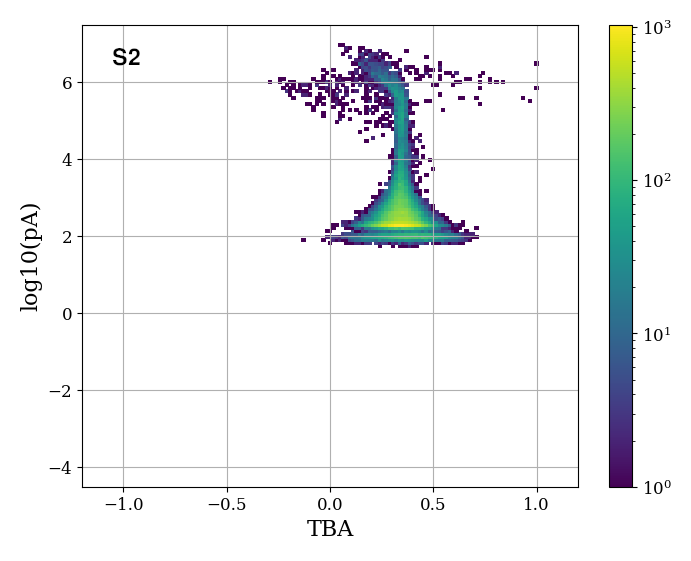}
   \includegraphics[width=0.45\textwidth]{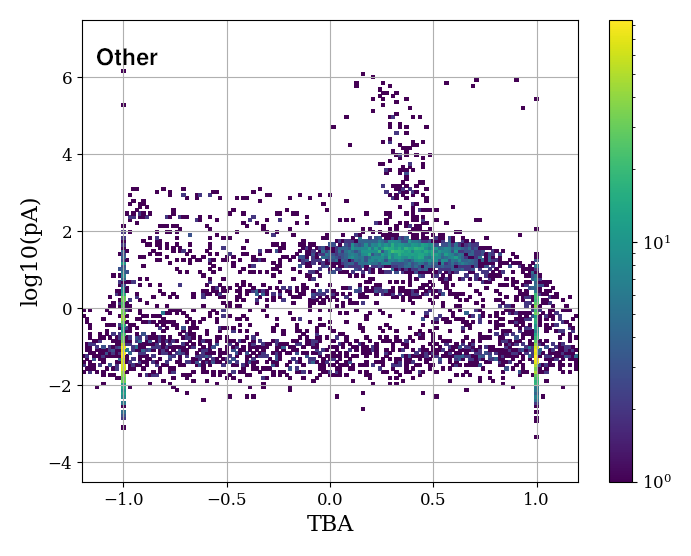}
   \includegraphics[width=0.45\textwidth]{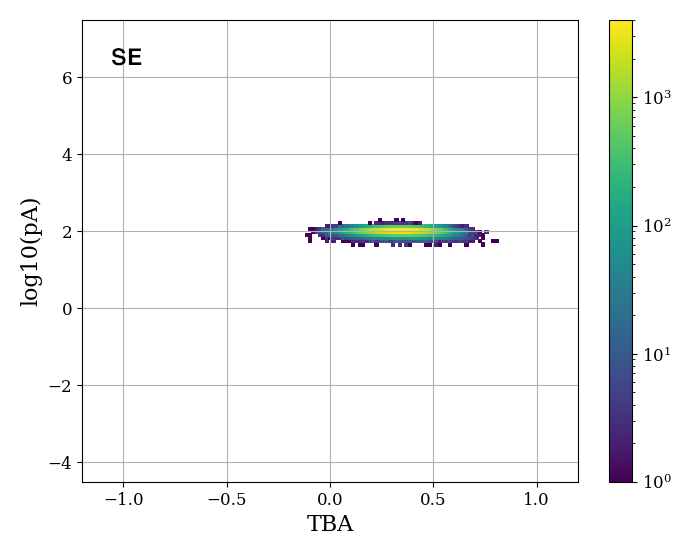}
   \caption{Distribution of the pulse populations in the marginal distribution \textit{TBA} vs \textit{pA} after being processed by the RFClassifier. The top-left plot displays the distribution of pulses classified as S1. The top right plot displays the distribution of S2 pulses, while the bottom plots display the Other population on the left and the SE population on the right.}
   \label{fig:RF_GMM_results_TBApA}
\end{figure*}
\begin{figure*}[!ht]
   \centering
   \includegraphics[width=0.45\textwidth]{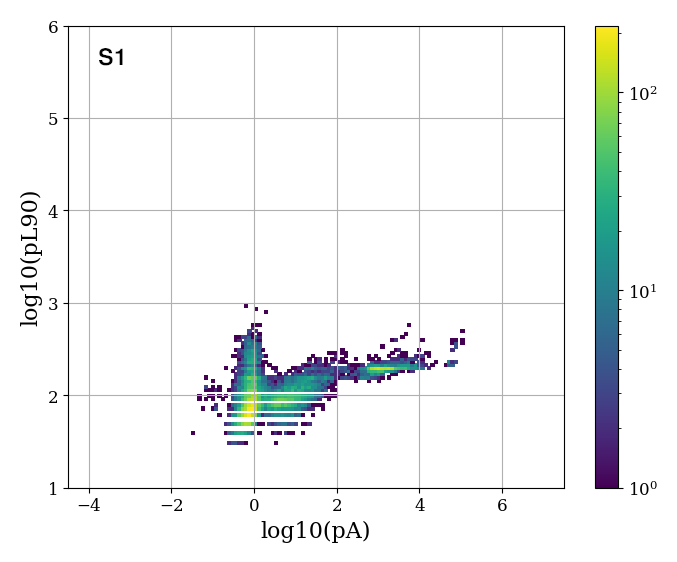}
   \includegraphics[width=0.45\textwidth]{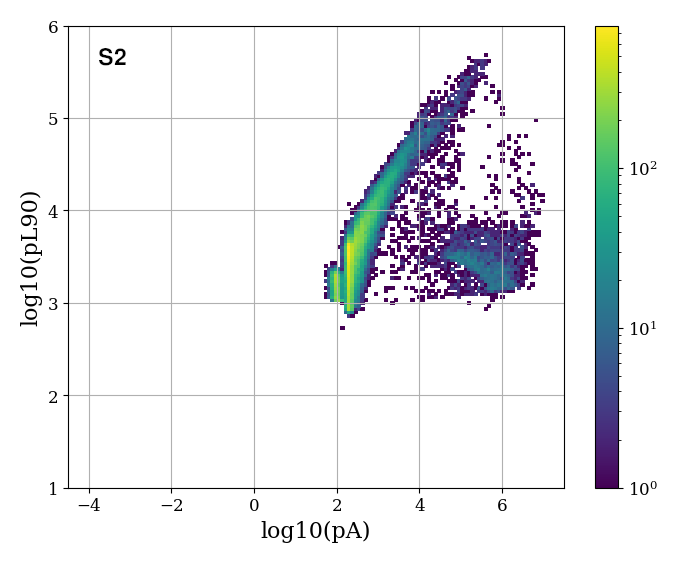}
   \includegraphics[width=0.45\textwidth]{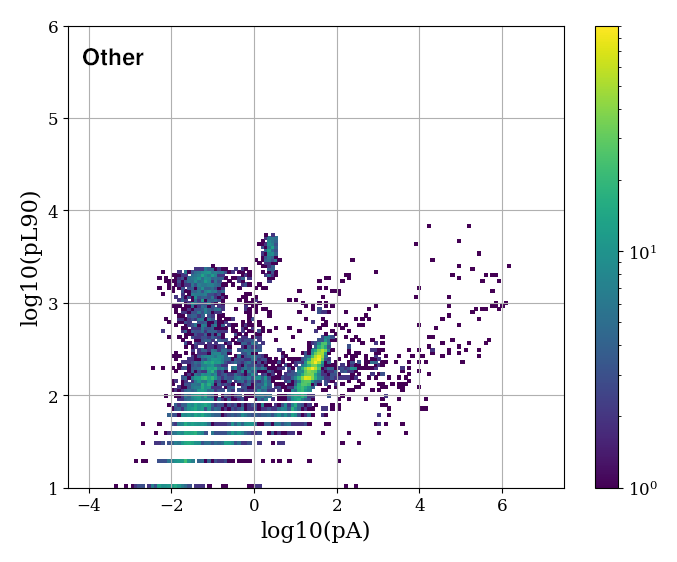}
   \includegraphics[width=0.45\textwidth]{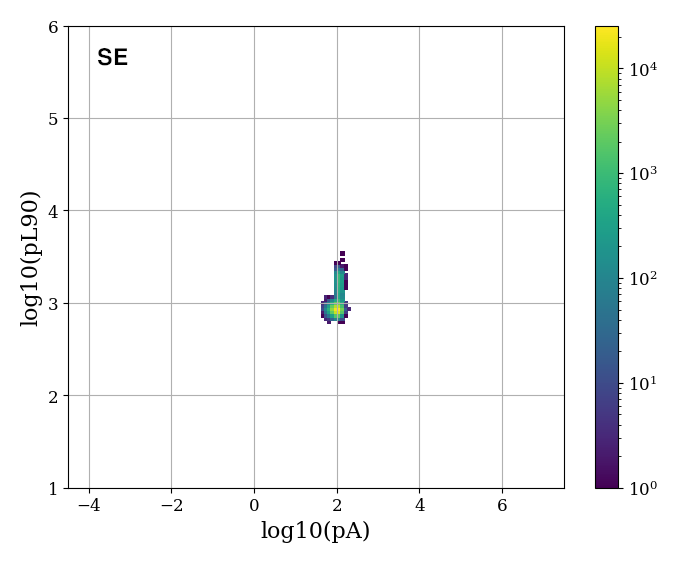}
   \caption{Distribution of the pulse populations in the marginal distribution \textit{pA} vs \textit{pL90} after being processed by the RFClassifier. The top-left plot displays the distribution of pulses classified as S1. The top right plot displays the distribution of S2 pulses, while the bottom plots display the Other population on the left and the SE population on the right.}
   \label{fig:RF_GMM_results_pApL90}
\end{figure*}

Figures \ref{fig:RF_GMM_results_TBApA} and \ref{fig:RF_GMM_results_pApL90} display the results of the predictions of the RFClassifier model for the test dataset, represented in the marginal distributions $\log_{10}(\textit{pA})$ vs $\log_{10}(\textit{pL90})$ and \textit{TBA} vs $\log_{10}(\textit{pA})$, respectively. These distributions can be compared to the plots in Figure \ref{fig:LZap_populations}, that displays the distribution of all pulse populations conveniently labelled.
The population of SE split pulses is clearly visible at the center of the bottom-left plot of Figure \ref{fig:RF_GMM_results_TBApA}, that displays the distribution of the pulses classified as Other by the RFClassifier. This module successfully tagged most of the SE split pulses as Other, leaving only a small number of these spurious pulses in the distribution of pulses classified as S1, displayed on the top-left plot of Figure \ref{fig:RF_GMM_results_TBApA}.

Table \ref{tab:RF_GMM_confMat} displays the confusion matrix of the RFClassifier for the test set of the RQ dataset, using the GMM results as the target class. The overall validation accuracy (\textit{acc}) of the RFClassifier model, calculated using the GMM results as the target classification, is $\textit{acc} = 99.38\%$. 
Considering only the classification of S1-like and S2-like pulses, \ie, not considering the mixing of S2 and SE pulse labels to be a misclassification, the validation accuracy is $\textit{acc}^{\text{S1S2}} = 99.67\%$.
\begin{table*}[!ht]
    \centering
    \caption{Confusion matrix of the RFClassifier results over the test dataset, using the GMM results as the target labels.}
    \begin{tabular}{p{2.1cm} | p{1.5cm} p{1.4cm} p{1.4cm} p{1.4cm} | r r}
       \multicolumn{1}{c|}{} & \multicolumn{4}{c|}{\textbf{RFClassifier Predicted class}} & \\
       \hline
       \textbf{GMM class}	&	S1	&	S2	&	SE	&	Other	&	Total	& \\
       \hline
       S1	    &	11590	&	0	    &	0	    &	261	    &	11851	&	5.9\%\\
       S2		&	0	    &	51103	&	348	    &	4	    &	51455	&	25.7\% \\
       SE		&	0	    &	228	    &   128371	&	0	    &	128599	&	64.4\% \\
       Other	&	385	    &	13	    &	5	    &	7692	&	8095	&	4.0\% \\
       \hline
       \textbf{Total}	&	11975	&	51344	&	128724	&	7957	&	200000	&\\
    \end{tabular}
    \label{tab:RF_GMM_confMat}
\end{table*}

The main failure mode present in these results is the classification of Other pulses as S1-like pulses, mainly from SE splits. Conversely, some S1-like pulses are also being identified as Other pulses, hinting at some level of label mixing in the GMM results (see discussion in Section \ref{subsec:Discussion}).

\subsubsection{Feature importance ranking}
\label{subsubsec:FeatureImportance}

When working with classification of data with large dimensionality it is useful to rank the data features by their usefulness in partitioning the data. 
This allows ML models like those explored in this work to be trained with a smaller subset of features that ranked higher, and by doing so greatly improving training efficiency without compromising classification performance \cite{ElementsStatisticalLearning}, assuming that this feature ranking is ubiquitous across different models.
Even for data with lower dimensionality, the identification of the best discriminant features is an extremely useful step in the development of traditional heuristic classification algorithms, or in providing important information for the development of more efficient online data monitoring tools.

The permutation importance score method was used in this work to determine which RQs are the best overall discriminants for this dataset. This method evaluates the decrease in a model score when a single feature value is randomly shuffled, and is more reliable than the typical variable importance ranking methods based on impurity indices and variable frequency since it is less sensitive to highly correlated variables and to asymmetric representation of class labels \cite{RandomForestsBreiman,BioInformaticsBiasRFVariableImportance}.

\begin{figure}
   \centering
   \includegraphics[width=0.45\textwidth]{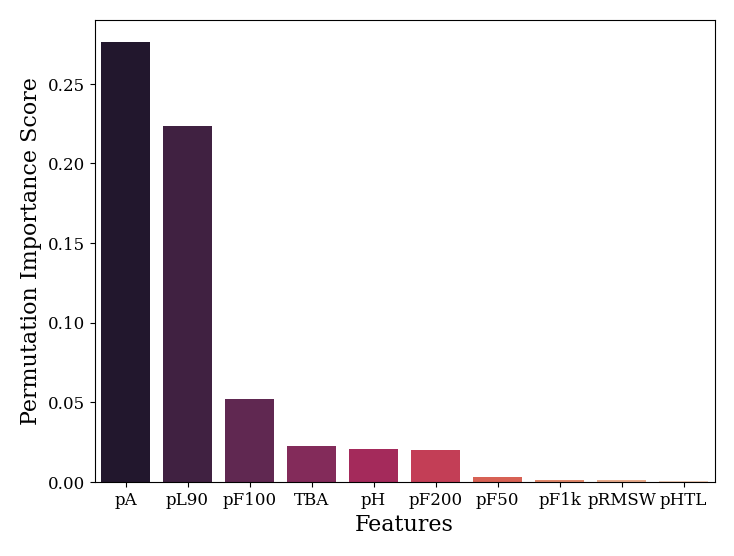}
   \caption{Permutation importance ranking obtained with the RFClassifier.}
   \label{fig:RF_PermImportance_barplots}
\end{figure}

Figure \ref{fig:RF_PermImportance_barplots} shows the permutation importance score for each RQ considered in the analysis. The scores were obtained using the fully trained RFClassifier model. 
The \textit{pA} and \textit{pL90} RQs are the best overall discriminants, with \textit{pF100}, \textit{pF200}, \textit{pH} and \textit{TBA} all roughly equal as the next strongest features. 
There is also an apparent preference for the \textit{pF100} RQ among the prompt fraction RQs. The \textit{pF50}, \textit{pF1k}, \textit{pRMSW} and \textit{pHTL} RQs scored the lowest on the permutation importance ranking. 

The feature ranking provided by the RFClassifier was used to inform the training of the NN ensemble method presented in Section \ref{subsec:TriNet}. 
The low dimensionality of the data used in this study, with only 10 independent features, means that training times are manageable and reducing the number of features will not impact it significantly. 
However, excluding the features with lower discrimination power increases the effectiveness of training and can help the model reach a higher accuracy. This was verified by comparing the accuracy of the same model trained with all features and with only the top 6 performing ones. 
For that reason, the \textit{pF50}, \textit{pF1k}, \textit{pRMSW} and \textit{pHTL} RQs were excluded during the training of the NN ensemble method presented in Section \ref{subsec:TriNet}.

These results were also used to inform other non-ML classification algorithms that were proposed for LZ.

%%%%%%%%%%%%%%%%%%%%%%%%%%%%%%%%%%%%%%%%%%%%%%%%%%%%%%%%%%%%%
\subsection{TriNet Classifier}
\label{subsec:TriNet}

It was established during this work that an ensemble of dense NNs as binary classifiers, each trained in a One-vs-All configuration, returned better overall results than a single multi-class NN classifier model. In a One-vs-All problem, each of the NNs focuses on learning how to distinguish a single class from the remaining.

The NN ensemble model developed in this work combines several independent learners, each trained separately to identify a single given categorical pulse class. The prediction of the ensemble of $K'$ NN classifiers is represented by a function $\hat{y} = f(\mathbf{x})$, with $\hat{y} = \{ \hat{y}_1 ; \hat{y}_2 ; \dotsc ; \hat{y}_{k'} ; \dotsc ; \hat{y}_{K'} \}$. The terms $\hat{y}_{k'}$ are the output vectors of the $k^{\text{th}}$ NN classifier in the ensemble, labelled here as $\text{NN}_{k'}$. These terms can be defined as independent functions such that $\hat{y}_{k'} = f_{k'}(\mathbf{x})$, with $f_{k'}$ being the function fitted by $\text{NN}_{k'}$. The terms $\hat{y}_{k'}$ are two-vectors of the form $\hat{y}_{k'} = \left( \epsilon_{k'} , \tilde{\epsilon}_{k'} \right)$. The component $\hat{y}_{k'}^1 = \epsilon_{k'}$ is the ``response'' of $\text{NN}_{k'}$, and should approach 1 when a pulse is of the class designated to the NN. Conversely, the term $\hat{y}_{k'}^2 = \tilde{\epsilon}_{k'}$ represents the response of $\text{NN}_{\text{k'}}$ for all the other classes, here named the ``anti-response''. Since the response and anti-response are anti-correlated, only the response is used in this work. 
The output of each NN is continuous and bounded, and can be viewed as the likelihood of a given pulse being from the respective target class. Therefore, this method provides a quasi-probabilistic approach to the classification problem. Obtaining probabilistic information instead of a categorical result can provide more information about the nature of the pulse and the degree of ambiguity of the prediction. 

In this work the Other pulses are considered the ``exception'' class to the remaining target pulse classes and thus they do not need to be learned explicitly by the NN ensemble model. For that reason, the number of target classes with a dedicated NN classifier in the ensemble is $K'=3$, those classes being S1, S2 and SE, as mentioned at the end of Section \ref{sec:method}. Therefore, this model was named \textit{TriNet Classifier}\footnote{\url{https://gitlab.com/PauloBras/trinetclassifier.git}}\footnote{\url{https://gitlab.com/PauloBras/mlforpc.git}}, and was implemented and trained using the Keras library \cite{chollet2015keras} with a Tensorflow backend.

\begin{figure*}[!ht]
    \centering
    \includegraphics[width=0.6\textwidth]{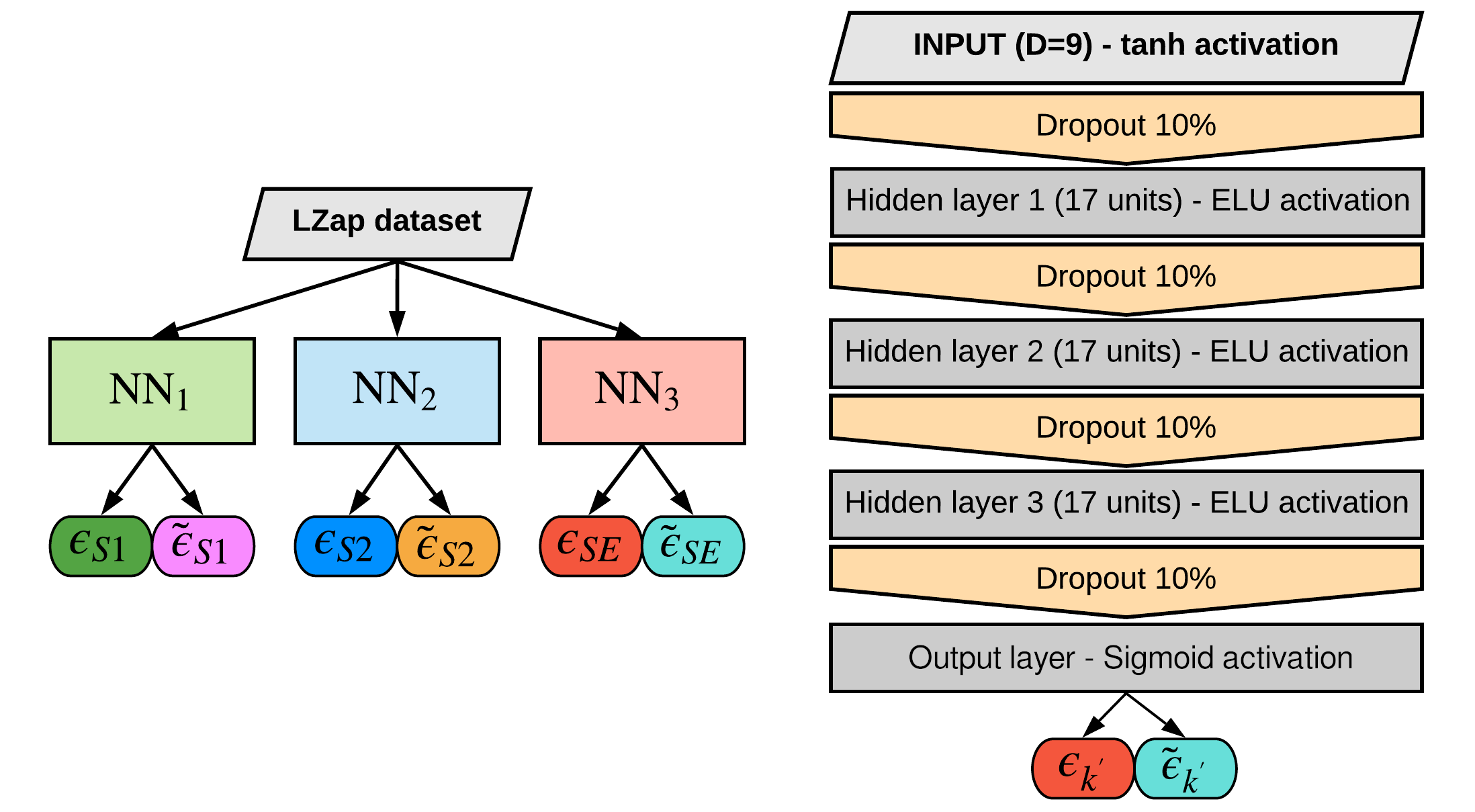}
    \caption{Simplified schematic of the TriNet Classifier model (left). The model is composed of an ensemble of three NN models (right) each trained to separate the three main pulse classes in the data: S1, S2 and SE pulses. The outputs of each NN classifier, $\hat{y}_{k'} = \{ \epsilon_{k'} , \tilde{\epsilon}_{k'} \}$, represent the response and anti-response of the classifier to its respective class.}
    \label{fig:TriNet_Diagram}
\end{figure*}

The architecture of the TriNet Classifier is represented in the left side of Figure \ref{fig:TriNet_Diagram}. The ensemble model is composed of three independently trained NN classifiers schematically depicted in the right side of Figure \ref{fig:TriNet_Diagram}. 
Each NN classifier has $L=3$ fully connected hidden layers, with $P=17$ hidden units (neurons) each for a total of $h=51$ neurons. These values were selected by monitoring the performance of the model with different architectures. 
All hidden neurons have an exponential linear activation (ELU) \cite{chollet2015keras}. The input layer and all hidden layers have an associated dropout layer that will randomly shut down neurons with a 10\% probability per neuron in each training batch in order to prevent overtraining.

The output of the TriNet Classifier is given by the ensemble of the individual outputs of each NN classifier, $\hat{y} = \{ \hat{y}_{1} ; \hat{y}_{2} ; \hat{y}_{3} \}$, with $\hat{y}_{k'} = \left( \epsilon_{k'} , \tilde{\epsilon}_{k'} \right)$ the output of the individual NN classifier assigned to class label $k'$. For simplicity, the output $\hat{y}$ is explicitly written as a matrix of the type
$$\hat{y} = \begin{bmatrix}
\epsilon_{1} & \tilde{\epsilon}_{1} \\
\epsilon_{2} & \tilde{\epsilon}_{2} \\
\epsilon_{3} & \tilde{\epsilon}_{3} \\
\end{bmatrix}$$
with the row vectors representing the individual outputs of the $\text{NN}_{k'}$ classifiers and the column vectors representing the overall response and anti-response of the ensemble, hereby defined by $\epsilon = \left( \epsilon_{1}, \epsilon_{2}, \epsilon_{3} \right)^T$ and $\tilde{\epsilon} = \left( \tilde{\epsilon}_{1}, \tilde{\epsilon}_{2}, \tilde{\epsilon}_{3} \right)^T$, respectively. Since the sum of the elements of each individual output $\hat{y}_{k'}$ is approximately unity, the sum of all elements of the TriNet Classifier output is approximately $K'$ (=3 in this work), \ie,
\begin{equation}
	\label{eq:TriNet_sumOfElements_K}
	\sum\limits_{k'=1}^{K'} \left( \epsilon_{k'} + \tilde{\epsilon}_{k'} \right) \approx K'.
\end{equation}

For each of the pulse classes present in the data, the output $\hat{y}$ of the TriNet Classifier ensemble is expected to be asymptotically equivalent to
\begin{gather}
    f(\mathbf{x}|y=1) \sim
    \begin{bmatrix}
        1 & 0 \\
        0 & 1 \\
        0 & 1 \\
    \end{bmatrix} \: ;  \notag\\
    f(\mathbf{x}|y=2) \sim
    \begin{bmatrix}
        0 & 1 \\
        1 & 0 \\
        0 & 1 \\
    \end{bmatrix} \: ; \\
    f(\mathbf{x}|y=3) \sim
    \begin{bmatrix}
        0 & 1 \\
        0 & 1 \\
        1 & 0 \\
    \end{bmatrix}, \notag
    \label{eq:TriNet_output_examples}
\end{gather}
with the training label $y=k'$ representing the pulse class $s_{k'} \in \{S1, S2, SE\}$.

The responses of the individual classifiers of the TriNet ensemble to a pulse that resembles neither an S1 nor an S2 nor an SE pulse, \ie, an Other pulse, are expected to be small, \ie, the output of the TriNet Classifier model associated to an Other pulses is expected to be asymptotically equivalent to
\begin{equation}
f(\mathbf{x}|y\neq\{1,2,3\}) \sim
\begin{bmatrix}
0 & 1 \\
0 & 1 \\
0 & 1 \\
\end{bmatrix}.
\label{eq:TriNet_output_Other}
\end{equation}

It is convenient to quantify the global strength of the responses $\epsilon$ and anti-responses $\tilde{\epsilon}$ in order to evaluate if there is any type of ambiguity on the predictions. The \textit{confidence} on the result, $\Gamma_{\epsilon}$, can be expressed by the sum of the elements of $\epsilon$, and an equivalent quantity can be defined for $\tilde{\epsilon}$, designated by $\Gamma_{\tilde{\epsilon}}$.
\begin{align}
	\label{eq:TriNet_confidence}
	\Gamma_{\epsilon} =& \sum\limits_{k'=1}^{K'} \epsilon_{k'} \quad \text{(confidence)} \\
	\Gamma_{\tilde{\epsilon}} =& \sum\limits_{k'=1}^{K'} \tilde{\epsilon}_{k'} \quad \text{(confidence complement)}
\end{align}
The parameter $\Gamma_{\tilde{\epsilon}}$ can be seen as the complement to the confidence score $\Gamma_{\epsilon}$, since the result in Equation \ref{eq:TriNet_sumOfElements_K} implies that $\Gamma_{\tilde{\epsilon}} \approx K' - \Gamma_{\epsilon}$.

% probabilistic vector
The output of the TriNet model can be converted into a probabilistic vector, $p(k)$, using a simple set of rules using the response vector $\epsilon$ and the confidence score $\Gamma_{\epsilon}$. Since there is no representation of spurious pulses in training, the probability vector should have $K'+1$ number of terms, one for each $K'$ primary class plus an additional term assigned to the remaining classes. In the case of the TriNet classifier, the $K'=3$ primary classes will return $K'+1 = 4$ probability results. 
For $K'=3$, the simplest parametrization of the output of the model is given by Equations \ref{eq:TriNet_results_noMix} and \ref{eq:TriNet_results_someMix}, for when the confidence $\Gamma_{\epsilon}$ is less than or equal to 1 or when its larger than 1, respectively. The latter case is dubbed "overconfidence" and automatically indicates that there is some conflict between the responses of the classifiers, which can be the result of some pulses having mixed properties or hinting towards problems in the data. 
\begin{enumerate}
    \item If $\Gamma_{\epsilon} \leq 1$:
    \begin{align}
        &p(k \in K') = \epsilon_{k'} \nonumber \\
        &p(k=4) = 1 - \Gamma_{\epsilon} \quad \text{(Other)}
        \label{eq:TriNet_results_noMix}
    \end{align}
    \item If $\Gamma_{\epsilon} > 1$ (overconfidence):
    \begin{align}
        &p(k=4) = \dfrac{1}{2} \left(\Gamma_{\epsilon} - 1 \right) \quad \text{(Other)} \nonumber \\
        &p(k \in K') = \epsilon_{k'} \dfrac{1-p(k=4)}{\Gamma_{\epsilon}}
        \label{eq:TriNet_results_someMix}
    \end{align}
\end{enumerate}

Using the information obtained by feature importance ranking with the RFClassifier (see Section \ref{subsubsec:FeatureImportance}), the following list of RQs were selected as input: \textit{pA}, \textit{pH}, \textit{pL90}, \textit{pF100}, \textit{pF200}, and \textit{TBA}.  
These inputs were normalized and mean-centered when necessary to cover similar ranges, in order to avoid loss of information and neuron ``death'' \cite{NNsAndDeepLearningAggarwal2018}. 
The mean-centering of the \textit{pA} and \textit{pL90} RQs is done with respect to the mean values for SE pulses since these are very well defined in the data and provide a natural middle point for the S1-like and S2-like phase-spaces. 
The \textit{pH}, \textit{pA} and \textit{pL90} RQs were projected to a logarithmic representation since they span several orders of magnitude. The \textit{TBA} RQ and the prompt fraction family of RQs do not need normalization.

The dataset was divided into a training set and test set with 80--20\% splitting ratio. The training is optimized using the \textit{RMSprop} algorithm \cite{chollet2015keras} with an initial learning rate $\alpha=0.001$ and batch size $n=128$ samples. The choice of the hyperparameters of the model was performed \textit{a-priori} by monitoring its generalization accuracy, loss over the generalization data and average training time. The training is monitored using the accuracy score of each model over the validation dataset, with an early stopping of the training if the validation accuracy does not improve for 10 consecutive epochs. To avoid overfitting, the model is saved at the state it was at the beginning of the 10 final training epochs.

The class labels used to train the TriNet Classifier were obtained using the GMM clustering analysis explained in Section \ref{subsec:MLClustering}. As mentioned before, there is some degree of mixture between SE split pulses and S1 afterpulsing in the GMM results, which is expected to influence the performance of the TriNet Classifier in some way. If the mixture is not too severe, each NN is expected to overcome the label impurity under the \textit{Classification Without Labels} (CWoLA) paradigm \cite{Metodiev:2017vrx}. However, the contamination of SPE pulses with \textit{coincidence} larger than 1 seems to be uniformly distributed across the S1 and Other pulse classes in the GMM results, which may lead to some confusion during training (see discussion below).

In order to represent the pulse populations for different classes obtained with the TriNet Classifier, the $\hat{y}$ output, and in turn the probabilistic vector of the pulse classes obtained using Equations \ref{eq:TriNet_results_noMix} and \ref{eq:TriNet_results_someMix}, are converted to categorical class labels corresponding to the 4 possible pulse classes considered in this work. Here, the categorical classifications are obtained by setting threshold values to the elements of the probabilistic vector, with each threshold being tuned to minimize the misclassification errors (false positives and false negatives) for its respective class. This conversion step can be achieved in many different ways, and the one used in this work is not assumed to be optimal.

\begin{figure*}[!ht]
    \centering
    \includegraphics[width=0.45\textwidth]{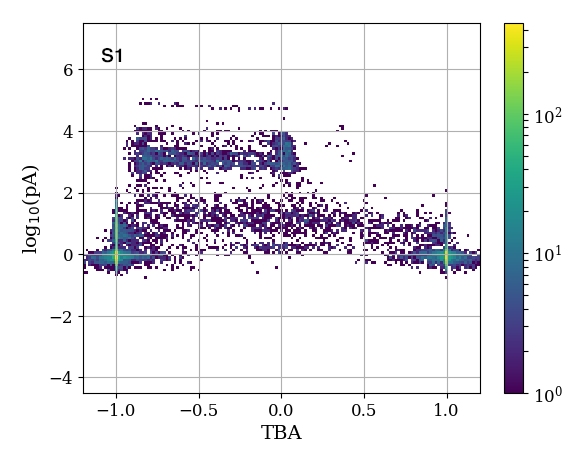}
    \includegraphics[width=0.45\textwidth]{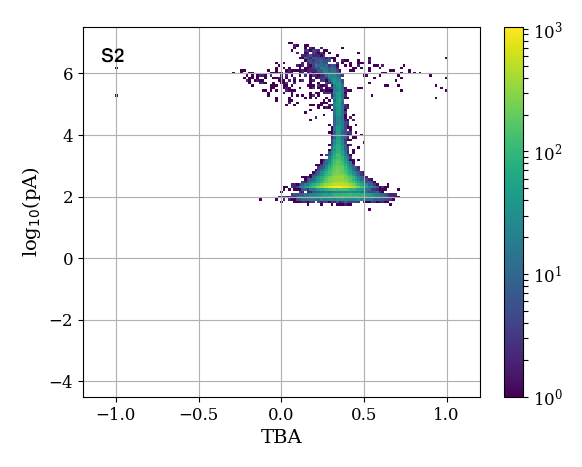}
    \includegraphics[width=0.45\textwidth]{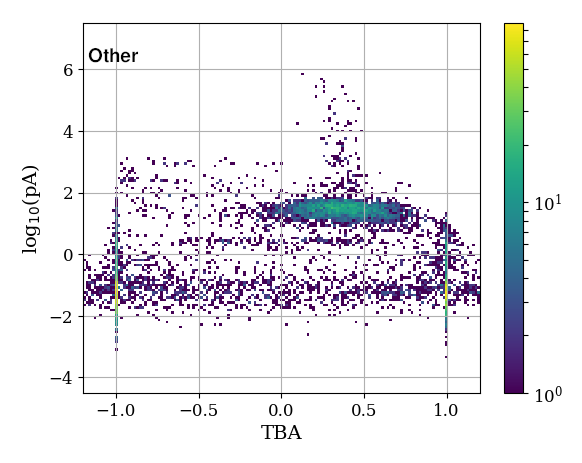}
    \includegraphics[width=0.45\textwidth]{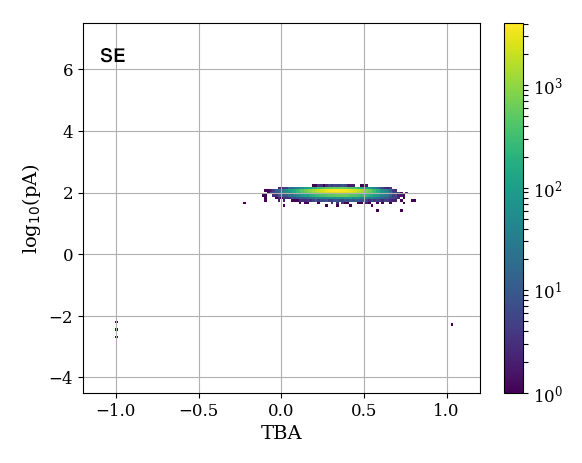}
    \caption{Distribution of the pulse populations in the marginal distribution \textit{TBA} vs \textit{pA} after being processed by the TriNet Classifier trained with the GMM results, and converted to categorical classifications. Top-left: pulses classified as S1. Top right: pulses classified as S2. Bottom-right: pulses classified as SE. Bottom-left: distribution of the remaining pulses, \ie, classified as Other.}
    \label{fig:TriNet_GMM_results_TBApA}
\end{figure*}

\begin{figure*}[!ht]
    \centering
    \includegraphics[width=0.45\textwidth]{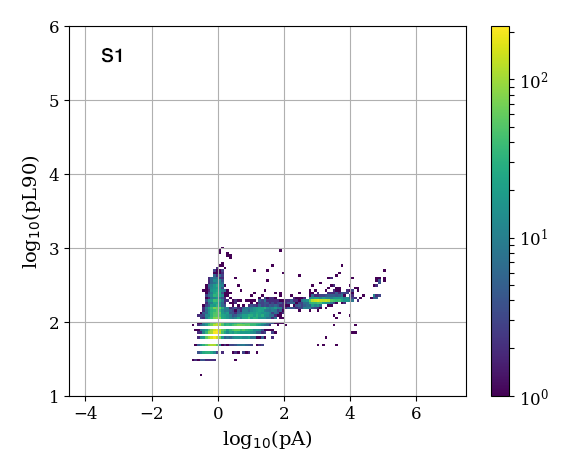}
    \includegraphics[width=0.45\textwidth]{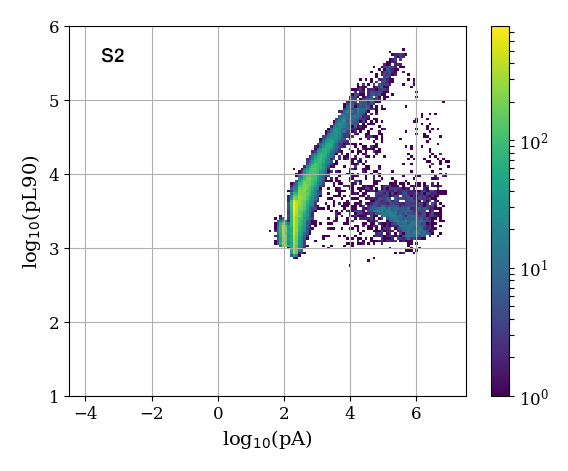}
    \includegraphics[width=0.45\textwidth]{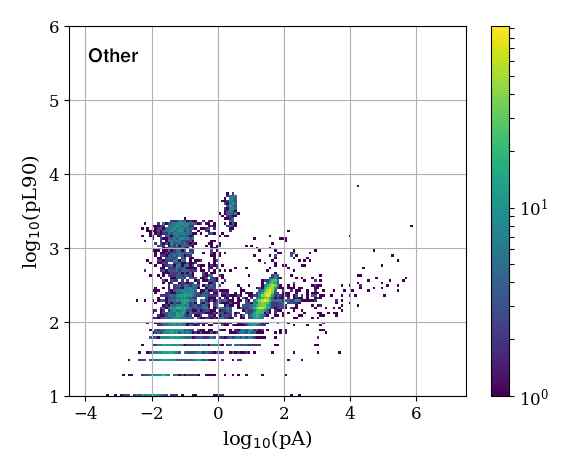}
    \includegraphics[width=0.45\textwidth]{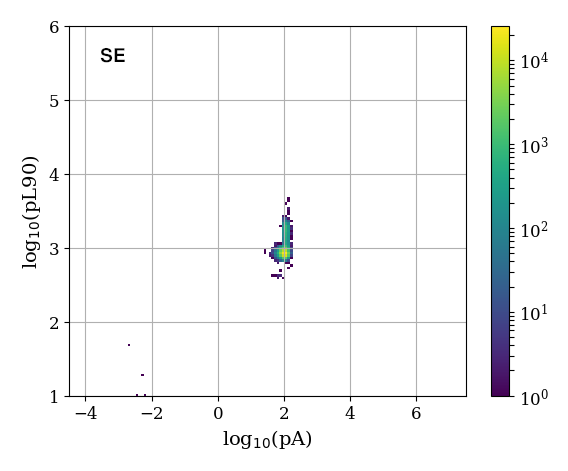}
    \caption{Distribution of the pulse populations in the marginal distribution \textit{pA} vs \textit{pL90} after being processed by the TriNet Classifier trained with the GMM results, and converted to categorical classifications. Top-left: pulses classified as S1. Top right: pulses classified as S2. Bottom-right: pulses classified as SE. Bottom-left: distribution of the remaining pulses, \ie, classified as Other.}
    \label{fig:TriNet_GMM_results_pApL}
\end{figure*}

Figures \ref{fig:TriNet_GMM_results_TBApA} and \ref{fig:TriNet_GMM_results_pApL} display the results of the predictions of the TriNet classifier model for the test dataset after converting them into categorical classifications, represented in the marginal distributions $\log_{10}(\textit{pA})$ vs $\log_{10}(\textit{pL90})$ and \textit{TBA} vs $\log_{10}(\textit{pA})$, respectively. These distributions can be compared to the plots in Figure \ref{fig:LZap_populations}, as well as with the results obtained with the RFClassifier model in Figures \ref{fig:RF_GMM_results_TBApA} and \ref{fig:RF_GMM_results_pApL90}.
The TriNet model is able to separate the main pulses classes with high efficiency, comparable to the results from the RFClassifier model presented in Section \ref{subsec:DTClass}. Similarly to those results, the TriNet classifier model managed to separate the majority of the SE split pulses, clearly visible at the center of the bottom-left plot of Figure \ref{fig:TriNet_GMM_results_TBApA}, that displays the distribution of the pulses classified as Other. The classification of some SE pulses with larger pulse length as S2 pulses is a consequence of both models being trained with the results from the GMM clustering analysis, discussed in Section \ref{subsec:MLClustering}. The sub-population of SPE pulses with \textit{coincidence} RQ larger than 2 (easily identified by having $\textit{TBA}\approx\pm1$) is present in both the distribution of pulses classified as S1 (top left plot of Figure \ref{fig:TriNet_GMM_results_TBApA}) and in the distribution of pulses classified as Other (bottom left plot of Figure \ref{fig:TriNet_GMM_results_TBApA}). This indicates that some mixing in the classification is taking place as mentioned before. A more detailed discussion is presented in Section \ref{subsec:Discussion}. 
Finally, it has been demonstrated that the TriNet model is able to identify spurious pulses without having explicitly learned from examples of these pulses during training, which may provide a strong method to identify unexpected pathological behaviour in the data by looking at pulses classified as Other.

Table \ref{tab:TriNet_GMM_vs_GMM_confMat} compares the results of the TriNet Classifier with the labels obtained in the GMM clustering analysis developed in Section \ref{subsec:MLClustering}.
\begin{table*}[!ht]
    \centering
    \caption{Confusion matrix of the results from the TriNet Classifier, trained with the GMM results as labels, compared with the GMM results obtained in Section \ref{subsec:MLClustering}.}
    \begin{tabular}{p{2.1cm} | p{1.5cm} p{1.4cm} p{1.4cm} p{1.4cm} | r r}
       \multicolumn{1}{c|}{} & \multicolumn{4}{c|}{\textbf{TriNet Predicted class}} & \\
       \hline
       \textbf{GMM class}	&	S1	&	S2	&	SE	&	Other	&	Total	& \\
       \hline
       S1	    &	11571	&	0	    &	0	    &	280	    &	11851	&	5.9\%\\
       S2		&	0	    &	51001	&	444	    &	10	    &	51455	&	25.7\% \\
       SE		&	0	    &	380	    &	128211	&	8	    &	128599	&	64.4\% \\
       Other	&	698	    &	38	    &	28	    &	7331	&	8095	&	4.0\% \\
       \hline
       \textbf{Total}	&	12269	&	51419	&	128683	&	7629	&	200000	&\\
    \end{tabular}
    \label{tab:TriNet_GMM_vs_GMM_confMat}
\end{table*}
From table \ref{tab:TriNet_GMM_vs_GMM_confMat}, the overall validation accuracy of the TriNet Classifier model, calculated over the GMM labels, is estimated to be $\textit{acc} = 99.06\%$. 
Considering only the classification of S1-like and S2-like pulses, \ie, not considering the mixing of S2 and SE pulse labels to be a misclassification, the validation accuracy becomes $\textit{acc}^{\text{S1S2}} = 99.47\%$. Similarly to the RFClassifier results in Section \ref{subsec:DTClass}, the main contribution to the accuracy loss of the TriNet model is the confusion between S1 pulses and Other pulses. This issue is discussed in detail in the next section.

%%%%%%%%%%%%%%%%%%%%%%%%%%%%%%%%%%%%%%%%%%%%%%%%%%%%%%%%%%%%%
\subsection{Discussion}
\label{subsec:Discussion}

The RFClassifier and the TriNet models show some tension with the results from the GMM clustering, especially in the mixing of S1 and Other labels. 
A detailed handscan focused on these cases was performed in order to determine their cause. 
This exercise indicated that the major contributions to the loss of accuracy on both the TriNet and RFClassifier models are the misclassification of SPE pulses with incorrect channel coincidence values, and a strong confusion between S1 afterpulsing and the SE split pulses.

Roughly 55\% of the pulses labelled as S1 in the GMM analysis, and 43\% of those labelled as Other, have coincidence lower than 3. The majority of these cases are SPE pulses with miscalculated coincidence, with a minor contribution from random noise (baselines). From the handscan, it was determined that these cases contribute to 64\% of the pulses misclassified as S1 and 24\% of those misclassified as Other by the TriNet model. If these pulses are excluded, the accuracy of the TriNet model increases up to 99.3\%.

The confusion between S1 afterpulsing and the SE split pulses seems to be caused by some mixture on the GMM analysis results, which influenced the TriNet and RFClassifier models. The degree of confusion indicates that the label mixture for these two pulse types is high, as mentioned previously in Section \ref{subsec:MLClustering}, which most likely prevented the models from fully learning to distinguish these two pulse types \cite{Metodiev:2017vrx}. Very often these pulses share the same RQ phase-space and have very similar waveforms. In fact, it became apparent during the handscan that discriminating by eye between some SE split pulses and S1 afterpulsing was remarkably difficult without some context from the rest of the event. Considering that the classification is performed at the pulse level without any information from the rest of the event (and only in the RQ space) it seems reasonable that the GMM analysis, and consequently the predictive models, could not fully separate these two populations.

The accuracy of tagging the primary S1 and S2 pulses correctly, those pulses that are essential to fully describe an event, is much higher than the values of the global accuracy presented here. Most of the classification issues mentioned above only involve pulses that are not critical to the physics analysis and are not expected to impact the performance of the detector in any significant way.

%%%%%%%%%%%%%%%%%%%%%%%%%%%%%%%%%%%%%%%%%%%%%%%%%%%%%%%%%%%%%
\section{Conclusions}
\label{sec:concl}

The methodology presented here demonstrates how one can develop new or improve existing algorithms aimed at the classification of signals in dual-phase noble element TPCs using standard ML methods. 
Even though this work was performed using simulated data, the omission of truth information and the realism of the simulations used ensure that this data is analogous to real detector data, which indicates that this analysis can accomplish the same results and overcome similar challenges when applied to real data. 

The initial clustering analysis and the feature importance ranking methods presented here provide vital information, directly from detector data, that can then be used to train dedicated ML predictive models capable of competing with common \textit{ad hoc} and heuristic methods. 
The GMM clustering analysis explored here provides a robust and minimally biased way of characterizing detector data directly, in order to train pulse classification algorithms while avoiding any dependencies on simulations. This technique will be tested with real detector data once LZ starts data taking. 

The RFClassifier model achieved an overall classification accuracy of 99.38\% and was able to successfully tag known spurious pulses in the data. 
The TriNet model also achieved an excellent global accuracy of 99.04\% and demonstrated that it is possible to teach an ensemble of NNs to identify spurious pulses while maintaining a high global classification accuracy, even if the spurious pulses are not explicitly learned. 
Both models perform at least on par with conventional methods as standalone pulse classification tools \cite{Aprile:2018dbl,Wang:2020coa,Akerib:2013tjd}, having both achieved a global accuracy $>99\%$. It was also demonstrated that these models can be used as auxiliary tools to identify systemic issues in the data processing stages preceding the classification. The TriNet Classifier model is also able to deliver probabilistic information, providing this classifier with a higher classification flexibility than the remaining methods studied here.

The results presented in this work were obtained whilst training the classification models with impure data. This contamination of the training labels originated from the GMM clustering analysis, which in turn resulted from errors in preceding data handling stages, \eg, miscalculated \textit{coincidence} RQ or split pulses. 
These uncertainties in the data are to be expected during the first stages of development of a processing framework or when a detector begins taking data. 
Despite these challenges, the methods developed in this work outperformed traditional methods without resorting to \textit{a priori} knowledge of the data, such as truth information from simulations during validation. Thus, this methodology is able to effectively overcome these potential sources of bias. 
Furthermore, these methods can be used to diagnose potential problems in preceding data handling stages, which has been demonstrated in this work with some level of success. In particular the GMM clustering analysis has revealed issues with the \textit{TBA} RQ calculation by isolating the corresponding outlier population. This property can help to identify potential issues in real data as well.

Finally, besides the challenges contemplated in this work (mainly caused by data handling stages between data acquisition and processing) additional problems are to be expected when analysing real detector data (from LZ or otherwise). 
However, since this methodology uses a data-driven approach, it is expected to overcome any data-specific challenges with at least the same level of success as that of traditional methods, regardless of the nature of these challenges.

%%%%%%%%%%%%%%%%%%%%%%%%%%%%%%%%%%%%%%%%%%%%%%%%%%%%%%%%%%%%%
\section*{Acknowledgements}

The authors thank the LUX-ZEPLIN Collaboration for access to the simulated data used in this work, obtained using the BACCARAT simulation package, the Detector Electronics Response (DER) simulation package and the LZap data processing package. 
This work was supported by the Portuguese Foundation for Science and Technology (FCT) under the award numbers PTDC/FIS-PAR/2831/2020, and POCI/01-0145-FEDER-029147, PTDC/FIS-PAR/29147/2017, funded by OE/FCT, Lisboa2020, Compete2020, Portugal2020, FEDER. 
This work was also supported by a Ph.D. scholarship within the DAEPHYS doctoral program with award number PD/BD/114114/2015, funded by FCT.

\bibliography{bibliography}{}
\bibstyle{plain}

\end{document}